\documentclass[14pt]{revtex4}
 
 \usepackage{hyperref}
 \usepackage{amsmath,amsfonts,amssymb}
 \hypersetup{dvips,dvipdfm,colorlinks=true,urlcolor=magenta,filecolor=magenta,linktoc=page,citecolor=red,linkcolor=blue,bookmarks=true}
 \usepackage{graphicx,epstopdf}
\begin{document}
\title{ Variable $G$ and $\Lambda$ gravity theory and analytical Cosmological Solutions using Noether symmetry approach}

\author{Santu Mondal$^1$\footnote {santumondal050591@gmail.com}}
\author{Sourav Dutta$^2$\footnote {sduttaju@gmail.com}}
\author{Subenoy Chakraborty$^1$\footnote {schakraborty.math@gmail.com}}
\affiliation{$^1$Department of Mathematics, Jadavpur University, Kolkata-700032, West Bengal, India\\
		$^2$ Department of Pure Mathematics, University of calcutta, 35, Ballygunge Circular Rd, Ballygunge, Kolkata, West Bengal 700019}

\begin{abstract}
	
	The present work deals with scalar field cosmology in the framework of a quantum gravity modified Einstein-Hilbert Lagrangian with variable $G$ and $\Lambda$. Using Renormalization group, variable $G$ behaves as a minimally coupled filed (not the scalar-tensor theory) and variable $\Lambda$ can be interpreted as a potential function. The point Lagrangian for this model in the background of homogeneous and isotropic flat FLRW space-time model experiences point-like Noether symmetry and equivalent potential function $\Lambda(G)$ is determined. Using a point transformation in the $3D$ augmented space is found that one of the variable become cyclic and as a consequence there is considerable simplification to the physical system. Lastly, the constants of motion can be written in compact form and it is possible to have analytic cosmological solutions in the present context.\\

\end{abstract}

\maketitle
 Key words: Variable $G$; Variable $\Lambda$; Noether symmetry; Cosmological Solutions.

\section{Introduction}
The major challenge of modern cosmology is to accomodate recent observational evidences. A series of observational evidences based on the experimental study of (i) Supernova Ia (SNIa) \cite{n1, n2,  n3}, (ii) cosmic microwave background (CMB) radiation along with large -scale-structure surveys (LSS) \cite{n4, n5, n6}, (iii) baryon acoustic oscillation (BAO) \cite{n7, n8}, (iv) weak lensing \cite{n9} has shown that our Universe is going through a second accelerating phase of evolution and it is attributed to a peculiar matter source, the exotic matter termed as dark energy (DE).\\

The study of investigating unknown character of DE can be characterized in two families: in the framework of Einstein gravity a matter source having large negative pressure has been introduced \cite{n10, n11,  n12, n13, n14, n15, n16, n17, n18, n19, n20} (dark energy models) or alternatively changing the gravity sector i.e., modification of Einstein-Hilbert action (modified gravity theories) \cite{n21, n22, n23, n24, n25, n26, n27, n28, n29, n30, n31, n32, n33}. The present work is an attempt with the second proposal.\\

Usually, in many modified theories, the Newton's constant $G$ becomes a variable parameter (for example $G_{eff}=G\phi^{-1}$ in Brans-Dicke theory,  $G_{eff}=G f'(R)$ in $f(R)$-gravity theory) \cite{n21, n22, n34}. Also due to quantum fluctuations of the background metric \cite{n35, n36, n37, n38}, $G$ and $\Lambda$ are essentially spacetime functions. The behaviour of these variable parameters are governed by the renormalization group (RG) equations for a Wilson-type Einstein-Hilbert action where $\sqrt{g}$ is considered as an operator in the neighbourhood of a non-perturbative ultraviolet fixed point \cite{n39}. In Weinberg sense \cite{n40} the theory is asymptotically safe i.e., non-perturbatively re-normalizable \cite{n41, n42, n43, n44}. It is to be noted that in cosmological context, there are various alternative gravity theories \cite{n38, n45, n46, n47, n48, n49, n50, n51, n52, n53, n54, n55, n56, n57, n58, n59} which are modified by the renormalization group.\\

Usually, geometrical symmetries (namely Lie point and Noether symmetries) related to space-time are very useful in solving/studying physical problems. In the context of Noether symmetry, the conserved charges are usually chosen as a selection criterion to discriminate similar physical processes \cite{n60, n61, n62, n63, n64, n65, n66, n67}. Further, from mathematical point of view, the first integral or Noether integral related to Noether symmetry  provides a tool to simplify a given system of differential equation or to determine the integrability of the system. Moreover, one can constrain the physical parameters involved in a physical system by imposing the symmetry \cite{n68}. It is worthy to mention that in recent years there are lot of works \cite{n69, n70, n71, n72, n73, n74, n75} related  to the above symmetries in physical system in Riemannian spaces, the present work is an example where Noether symmetry is applied to a physical system in cosmological context having dynamical $G$ and $\Lambda$. Here the variable $\Lambda$  behaves as potential of a scalar field represented  by the dynamical $G$ and the potential function is determined from geometrical principle instead of a phenomenological choice. The evolution equations are simplified by finding a cyclic variable in the augmented space. Analytical solutions are determined and are analyzed from cosmological perspective. The paper is organized as follows: basic equation for the variable $G$, $\Lambda$ theory are presented for FLRW model in  section II. Section III deals with Noether symmetry approach and the potential function is determined from symmetry analysis. Cyclic variable in the augmented space is determined and simplified field equations are solved in section IV. Finally, cosmological implications to the analytical solutions are discussed from the point of view of recent observations are presented in section V.

 \section{Euler Lagrange Equation}
The action integral proposed by Bonanno et al.\cite{r1} on renormalization-group improving the modified ADM Lagrangian of General Relativity with variable G and $\Lambda$ cosmology can be expressed as 
\begin{eqnarray} 
S&=&S_m + \frac{1}{16\pi}\int \frac{N\sqrt{h}}{G}\Bigg(K_{ij}K^{ij}-K^2+R^*-2\Lambda G\Bigg)d^3x\nonumber\\
&+&\frac{\mu}{16\pi}\int\frac{N\sqrt{h}}{G}\bigg(N^{-2}(G_{,0})^2-2\frac{N^i}{N^2}G_{,0}G_{,i}-\bigg(h^{ij}-\frac{N^iN^j}{N^2}\bigg)G_{,i}G_{,j}\bigg)d^3x,\label{k1}
\end{eqnarray}
 where G and $\Lambda$ are functions of time, $\mu$ is the non-vanishing interaction term, $S_m$ is the matter source of the action integral and $K_{ij}$ represents the extrinsic curvature, the three dimensional metric tensor are presented by $h_{ij}$ with curvature $R^*$ and $N$ denotes the Lapse function.
 So, in Arnowitt-Deser-Misner(ADM) formalism the line element of the metric can be written as \cite{r2}, \cite{r3}
 
\begin{equation}
ds^2=-\bigg(N^2-N_iN^i\bigg)dt^2+2N_idtdx+h_{ij}dx^idx^j,~~~~ i,j=1,2,3. \label{k2}
\end{equation}
Now continue with the assumption of homogeneity and isotropy of the Universe the line element $(\ref{k2})$ becomes that for Friedmann-Lemaitre-Robertson-Walker(FLRW) Universe i.e.,

\begin{equation}
ds^2=-N^2dt^2+a^2(t)(dx^2+dy^2+dz^2).\label{k3}
\end{equation}
So, in the above configuration space $Q\big(N,a,G\big)$ the point like Lagrangian takes the form

\begin{equation}
\mathcal{L}(Q,\dot{Q})=\frac{1}{N}\bigg(-\frac{3}{G}a\dot{a}^2+\frac{\mu}{2G}a^3
\bigg(\frac{\dot{G}}{G}\bigg)^2\bigg)-Na^3V\big(G\big)+Na^3\rho_m,\label{k4}
\end{equation}
where $\Lambda(G)$=$G$$V(G)$, and $\rho_m=8\pi\rho_{m0}a^{-3(1+w)}$, represents the matter source of perfect fluid with $w$=$\frac{p_m}{\rho_{m}}$ as equation of state parameter and dot represents the derivative with respect to the time `$t$'.
The generalized second order Euler Lagrangian Equation takes the form
\begin{equation}
\frac{d}{dt}\bigg(\frac{\partial L}{\partial \dot{q_i}}\bigg)-\frac{\partial L}{\partial {q_i}}=0,\\ \label{k5}
\end{equation}
with $q_i$ as the generalized coordinate in the configuration space $Q=(q_i)$. The associated energy function 
\begin{equation}
E_L=\sum\dot{q_i}\frac{\partial L}{\partial \dot{q_i}}-L.\label{k6}
\end{equation}
In this problem variation with the metric variable $\bigg(N,a,G\bigg)$ to the point like Lagrangian(\ref{k4}) the second order Euler-Lagrange equation for $a$ and $G$ \cite{r1}, \cite{r4}, \cite{r5}                                                                                                                                                        
\begin{equation}
\ddot{a}+\frac{\dot{a}^2}{2a}-\frac{\dot{a}\dot{G}}{G}+\frac{\mu a\dot{G^2}}{4G^2}-\frac{1}{2}GaV-4w\pi\rho_{m_0}a^{-3w-2}=0,\label{k7}
\end{equation}
\begin{eqnarray}
\mu\ddot{G}+3\frac{\dot{a}\dot{G}}{a}-\frac{\dot{G}^2}{G}-3\frac{\dot{a}^2}{a^2G}+\frac{\mu\dot{G}^2}{G^2}+V(G)G=0,\label{k8}
\end{eqnarray}
and finally the total energy or the Hamiltonian constraint is given by 
\begin{equation}
-\frac{3}{G}a\dot{a}^2+\frac{\mu}{2G}a^3\bigg(\frac{\dot{G}}{G}\bigg)^2+a^3V(G)=8\pi\rho_{m_0}a^{-3w}.\label{k9}
\end{equation}
Here without any loss of generality one may assume that the lapse function to be constant.
For further procedure to solve the evolution equations (\ref{k7})-(\ref{k9}) the Noether symmetry approach is used in the next section.
\section{Noether symmetry approach}
According to Noether theorem a vector field on the tangent space TCS $\big(N,a,G,\dot{a},\dot{G}\big)$ \cite{r6, r7, r8}
\begin{equation}
X=\alpha\frac{\partial}{\partial a}+\beta\frac{\partial}{\partial G}+\gamma\frac{\partial}{\partial N}+\dot{\alpha}\frac{\partial}{\partial\dot{a}}+\dot{\beta}\frac{\partial}{\partial\dot{G}}, \label{k10}
\end{equation}
for which the Lie derivative of the Lagrangian (\ref{k4}) vanishes
\begin{equation}
\mathcal{L}_XL=0,\label{k11}
\end{equation}
is called the infinitesimal/symmetry vector for the system (i.e. there exists a Noether symmetry) and hence generates a conserved current.
Here~ $\alpha=\alpha\big(a,G\big)$, $\beta=\beta\big(a,G\big)$,   $\gamma=\gamma\big(a,G\big)$
are the functions in the configuration space and $\dot{\alpha}\big(a,G\big)=\frac{\partial\alpha}{\partial a}\dot{a}+\frac{\partial\alpha}{\partial G}\dot{G}$, $\dot{\beta}(a,G)=\frac{\partial\beta}{\partial a}\dot{a}+\frac{\partial\beta}{\partial G}\dot{G}$  with the total derivative operator is given by
\begin{equation}
D_t=\frac{\partial}{\partial t}+\dot{a}\frac{\partial}{\partial a}+\dot{G}\frac{\partial}{\partial G}.\label{k12} 
\end{equation} 
So, the conserved quantity associated with $X$ is given by 
\begin{equation}
I=\alpha(a,G)\frac{\partial L}{\partial\dot{a}} +\beta(a,G)\frac{\partial L}{\partial \dot{G}}.\label{k13}
\end{equation}
Now, from the Noether symmetry condition (\ref{k11}) we get the following set of partial differential equations
\begin{equation}
-\alpha+\frac{\beta a}{G}+\frac{\gamma a}{N}-2a\frac{\partial \alpha}{\partial a}=0\label{k14}
\end{equation}
\begin{equation}
-6\frac{\partial\alpha}{\partial G}+\frac{\mu a^2}{G^3}{\frac{\partial \beta
	}{\partial a}}=0\label{k15}
\end{equation}
\begin{equation}
3\alpha-\frac{3\beta a}{G}-\frac{\gamma a}{N}+2a\frac{\partial \beta}{\partial G}=0\label{k16}
\end{equation}
\begin{equation}
-3\alpha V(G)-24\pi w \rho_{m0} \alpha a^{-3(1+w)}-\beta a V -\frac{\gamma V(G)}{N}+\frac{8 \pi \rho_{m0} \gamma a^{-3w-2}}{N}=0.\label{k17}
\end{equation}
As $\alpha(a,G)$, $\beta(a,G)$, $\gamma(a,G)$ are the co-efficients of the symmetry vector $X$, so they must satisfy the over determined system of equations, in which from the first three set of equations (\ref{k14}-\ref{k16}) we get the explicit from of the $\alpha(a,G)$, $\beta(a,G)$, and $\gamma(a,G)$ using the method of separation of variable $\alpha=\alpha_1(a)\alpha_2(G)$, $\beta=\beta_1(a) \beta_2(G)$, $\gamma$=
$\gamma_1(a )\gamma_2(G)$, i.e.,
\begin{equation}
\alpha(a,G)=\alpha_0a^{p_1}G^{q_1},~ \beta(a,G)=\beta_0a^{p_1-1}G^{q_1+1},~\gamma(a,G)=
\gamma_0a^{p_1-1}G^{q_1},\label{k18}
\end{equation}
where $\alpha_0$, $\beta_0$, $\gamma_0$, $p_1$, $q_1$ are arbitrary parameters.
From equation (\ref{k17}) one has the restriction on the equation of state parameter as $w=-1$ and the potential is obtained as
\begin{equation}
V=8\pi \rho_{m0}-\frac{1}{c. G^{\frac{3\alpha_0 +\gamma
			_0}{\beta_0}}},\label{k19}
\end{equation}
with $c$ an integrating constant. Hence from equation(\ref{k13}) we get the corresponding conserved current
\begin{equation}
I=\alpha_0a^{p_1}G^{q_1}\frac{1}{N}\bigg(-\frac{6}{G}a\dot{a}\bigg)+\frac{1}{N}\beta_0a^{p_1-1}G^{q_1+1}.\bigg({\mu}a^3
\frac{\dot{G}}{G^3}\bigg).\label{k20}
\end{equation}
\section{The exact solution }
In this section we shall turn our attention to find the exact cosmological solution of the evolution equations (\ref{k7}) and (\ref{k8}). To simplify the evolution equations we need the transformation of co-ordinates $(a,G,N)$ $\rightarrow$ $(u,v,W)$ such that inner product operator with $X$ results \cite{r9, r10, r11, r12, r13, r14}
\begin{equation}
i_{\overrightarrow{X}}du=1,~~i_{\overrightarrow{X}}dv=0,~~i_{\overrightarrow{X}}dW=0,\label{k21}
\end{equation}
i.e
\begin{eqnarray}
\alpha\frac{\partial u(a,G)}{\partial a}+\beta\frac{\partial u(a,G)}{\partial G}+\gamma\frac{\partial u(a,G)}{\partial N}=1\nonumber\\
\alpha\frac{\partial v(a,G)}{\partial a}+\beta\frac{\partial v(a,G)}{\partial G}+\gamma\frac{\partial v(a,G)}{\partial N}=0\label{k22}\\
\alpha\frac{\partial W(a,G)}{\partial a }+\beta\frac{\partial W(a,G)}{\partial G}+\gamma\frac{\partial W(a,G)}{\partial N}=0,\nonumber
\end{eqnarray}
and as a consequence $u$ is a cyclic co-ordinate. Due to the above  transformation the symmetry vector $\overrightarrow{X}$ changes to
\begin{equation}
{\tilde{X}}=(i_{\overrightarrow{X}}du)\frac{\partial}{\partial u}+(i_{\overrightarrow{X}}dv)\frac{\partial}{\partial v}+(i_{\overrightarrow{X}}dw)\frac{\partial}{\partial w}+(\frac{d}{dt}(i_{\overrightarrow{X}}du))\frac{d}{d\dot u}+(\frac{d}{dt}(i_{\overrightarrow{X}}dv))\frac{d}{d\dot v}+(\frac{d}{dt}(i_{\overrightarrow{X}}dW))\frac{d}{d\dot{W}}\label{k23}.
\end{equation}
Solving this set of equations (\ref{k22}) one can get 
\begin{equation}
a=e^u,~\frac{a}{G}=e^v,~\frac{a^3}{NG}=e^W,\label{k24}
\end{equation}
As a consequence the transformed energy function and the transformed point like Lagrangian(\ref{k4}) in the new coordinates look like
\begin{equation}
E= e^{W}\bigg[\bigg(9+\frac{\mu}{2}\bigg)\dot{u}^2-
\mu\dot{u}\dot{v}+\frac{\mu \dot{v}^2}{2}\bigg]-\frac{1}{c}e^{6v-W}  ,\label{k25}
\end{equation} 
\begin{equation}
\mathcal{L}(u,v,W,\dot{u},\dot{v})=e^{W}\bigg(-3\dot{u}^2+\frac{\mu\dot{u}^2}{2}-
\mu\dot{u}\dot{v}+\frac{\mu \dot{v}^2}{2}\bigg)+\frac{1}{c}e^{6v-W}.\label{k26}
\end{equation}
So the Euler-Lagrange equations in the new coordinates take the simple form
\begin{equation}
\ddot{u}=\frac{\mu}{\mu-6}\ddot{v},\label{k27}
\end{equation}
\begin{equation}
-\mu\ddot{u}+\mu\ddot{v}=\frac{6}{c}e^{6v-2W}.\label{k28}
\end{equation}  
  Now, eliminating $\ddot{u}$ from the equations (\ref{k27}) and (\ref{k28}) and integrating one has
\begin{equation}
\dot{v}^2=2\lambda\int e^{6v}dv+v_0,\label{k29}
\end{equation}
where $\lambda =\bigg(\frac{6-\mu}{c\mu}\bigg)e^{-2W}$ and $v_0$ is an integration constant.
Hence the explicit cosmological solutions are possible for the following cases\\
\begin{figure}
	\begin{minipage}{0.4\textwidth}
		\includegraphics[width=1\textwidth]{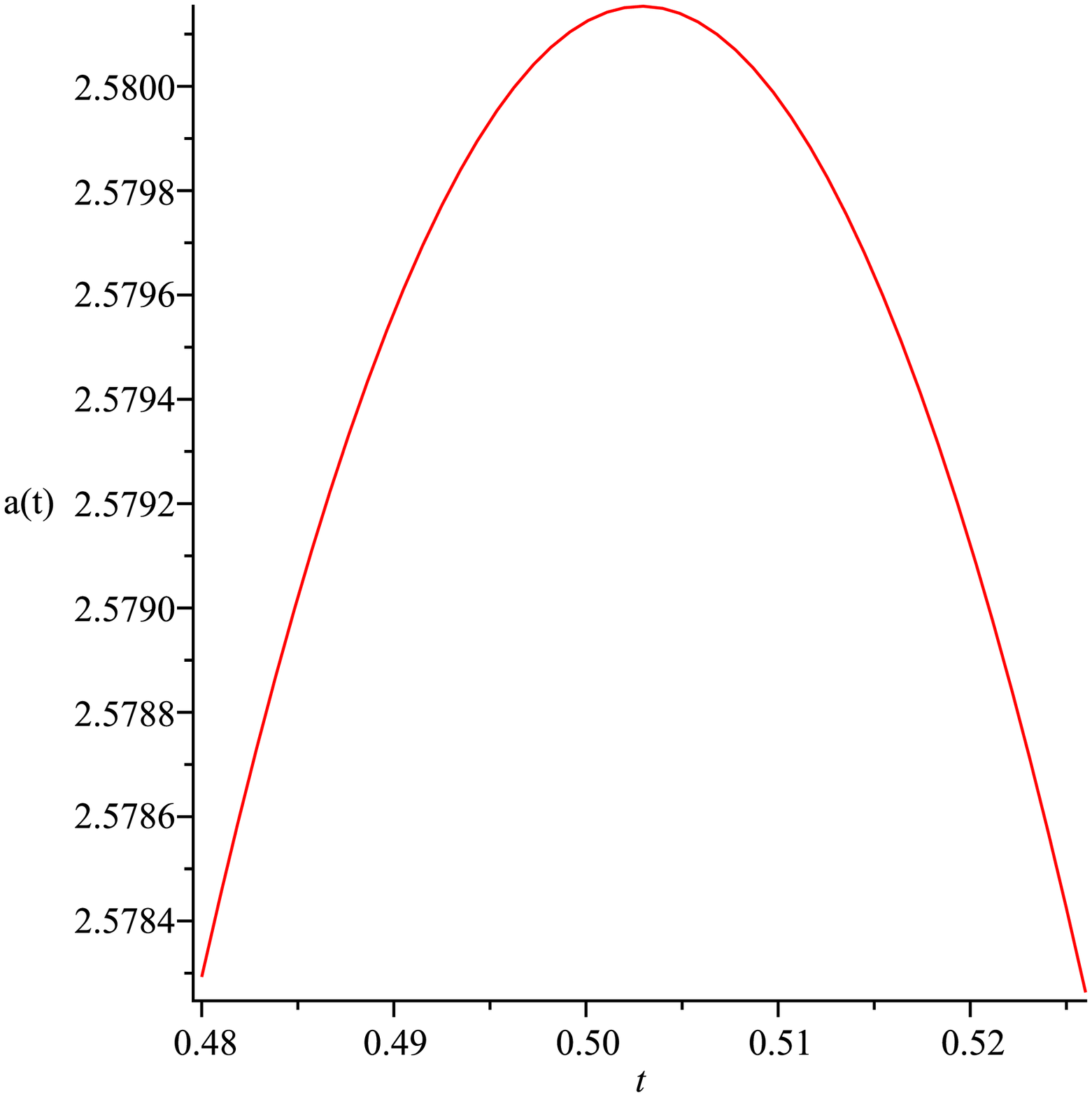}\\
		\caption{ The scale factor $a(t)$ is plotted against $t$~for equation (\ref{k32}) ($\lambda<0$,~ $v_0>0$). }
		\label{fig1}
	\end{minipage}
	\begin{minipage}{0.4\textwidth}
		\includegraphics[width=1\textwidth]{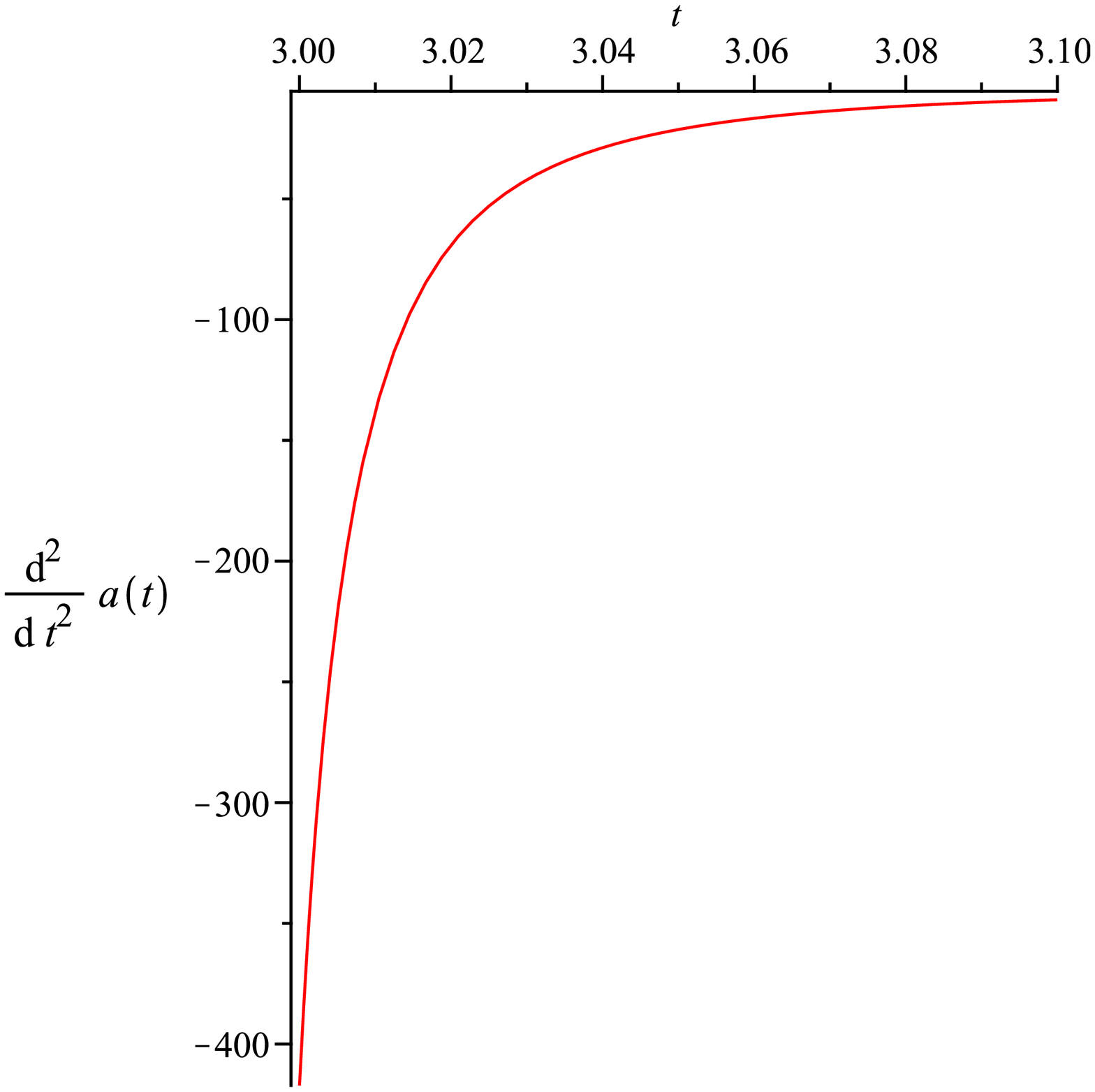}\\
		\caption{shows the evolution of the acceleration parameter with respect to cosmic time `$t$'~for equation  (\ref{k32}) ($\lambda<0$,~ $v_0>0$).}
		\label{fig2}
	\end{minipage}
\end{figure}

\begin{figure}
	\begin{minipage}{0.4\textwidth}
		\includegraphics[width=1\textwidth]{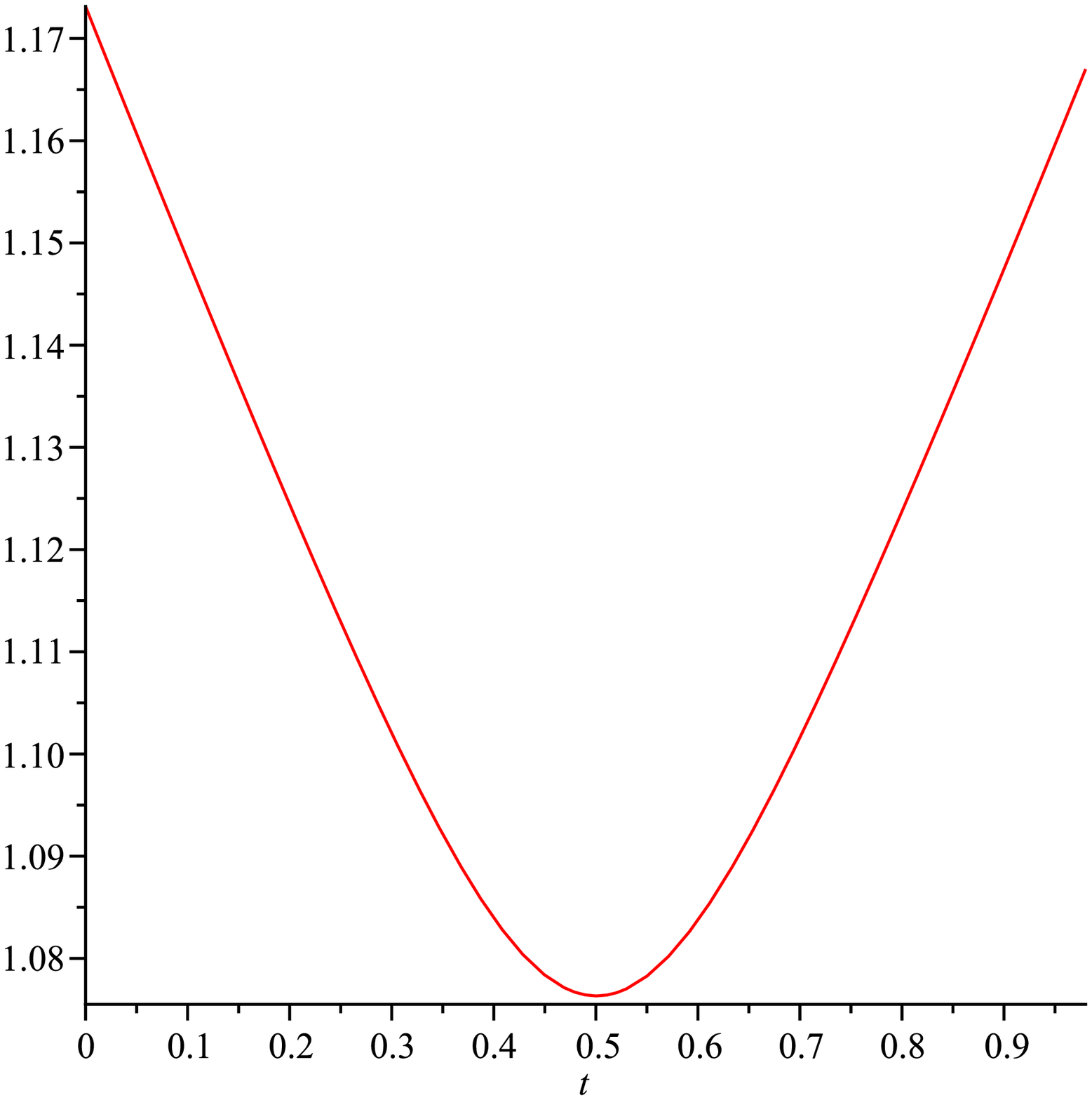}\\
		\caption{ Graphical representation of  $a(t)$ respect to cosmic time $t$~for equation (\ref{k33}) ($\lambda>0$,~ $v_0<0$). }
			\label{fig3}
		\end{minipage}
		\begin{minipage}{0.4\textwidth}
			\includegraphics[width=1\textwidth]{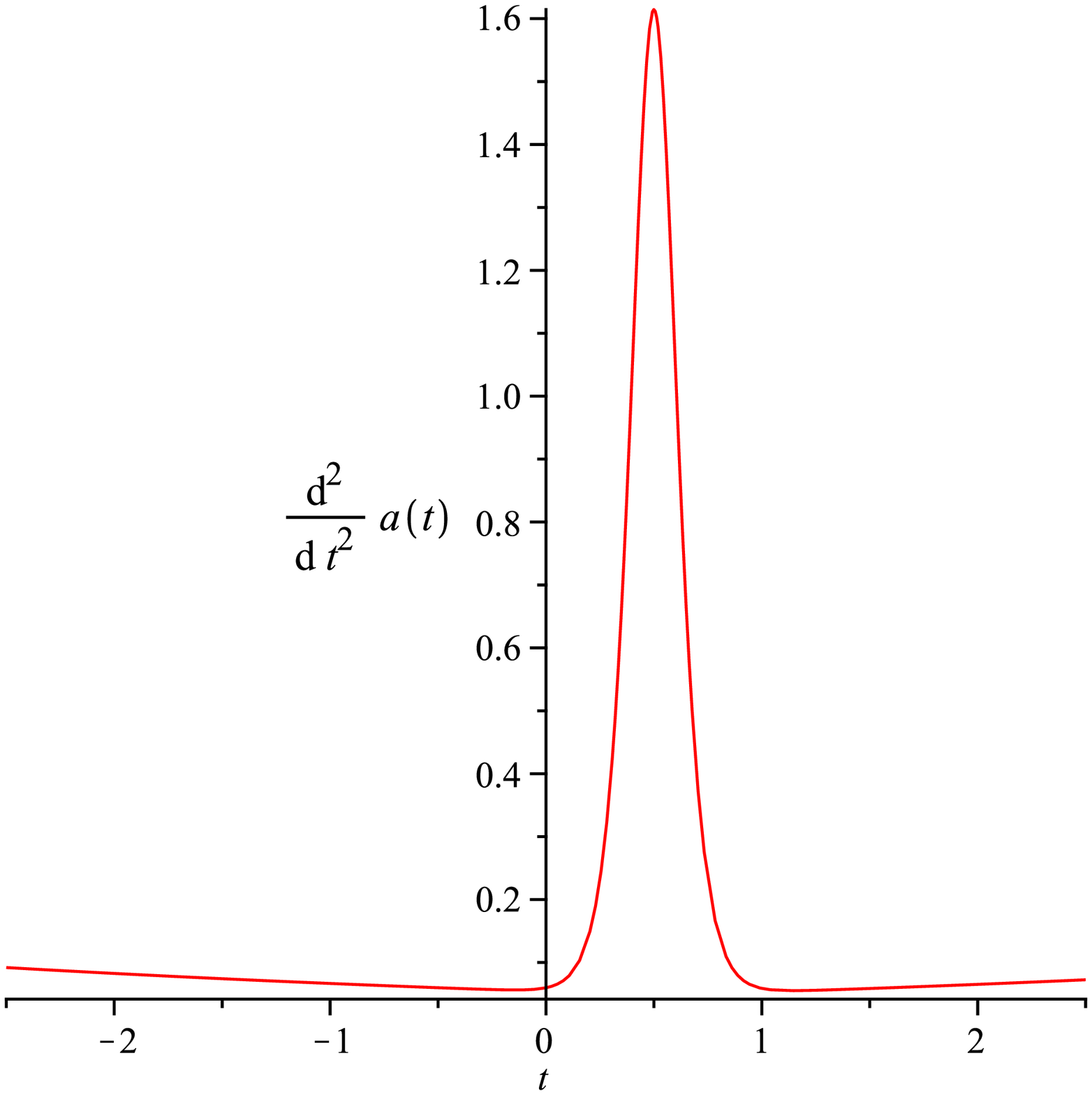}\\
			\caption{ Graphical representation of the acceleration parameter respect to cosmic time  $t$~for equation (\ref{k33}) ($\lambda>0$,~ $v_0<0$).}
			\label{fig4}
		\end{minipage}
	\end{figure}
	
	\begin{figure}
		\begin{minipage}{0.4\textwidth}
			\includegraphics[width=1\textwidth]{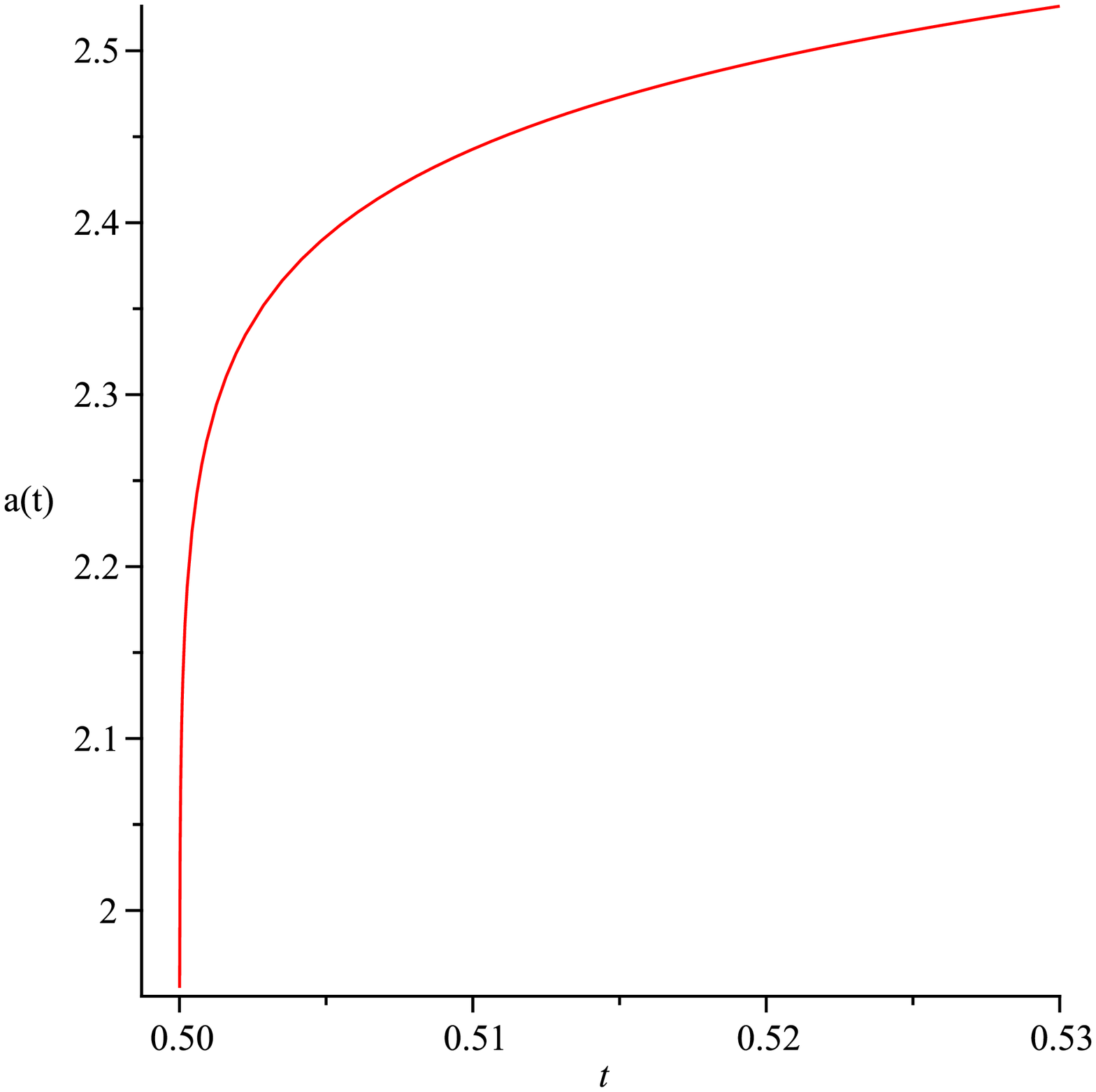}\\
			\caption{ The scale factor $a(t)$ is plotted against $t$~for equation  (\ref{k34}) ($\lambda>0$,~ $v_0>0$).}
			\label{fig5}
		\end{minipage}
		\begin{minipage}{0.4\textwidth}
			\includegraphics[width=1\textwidth]{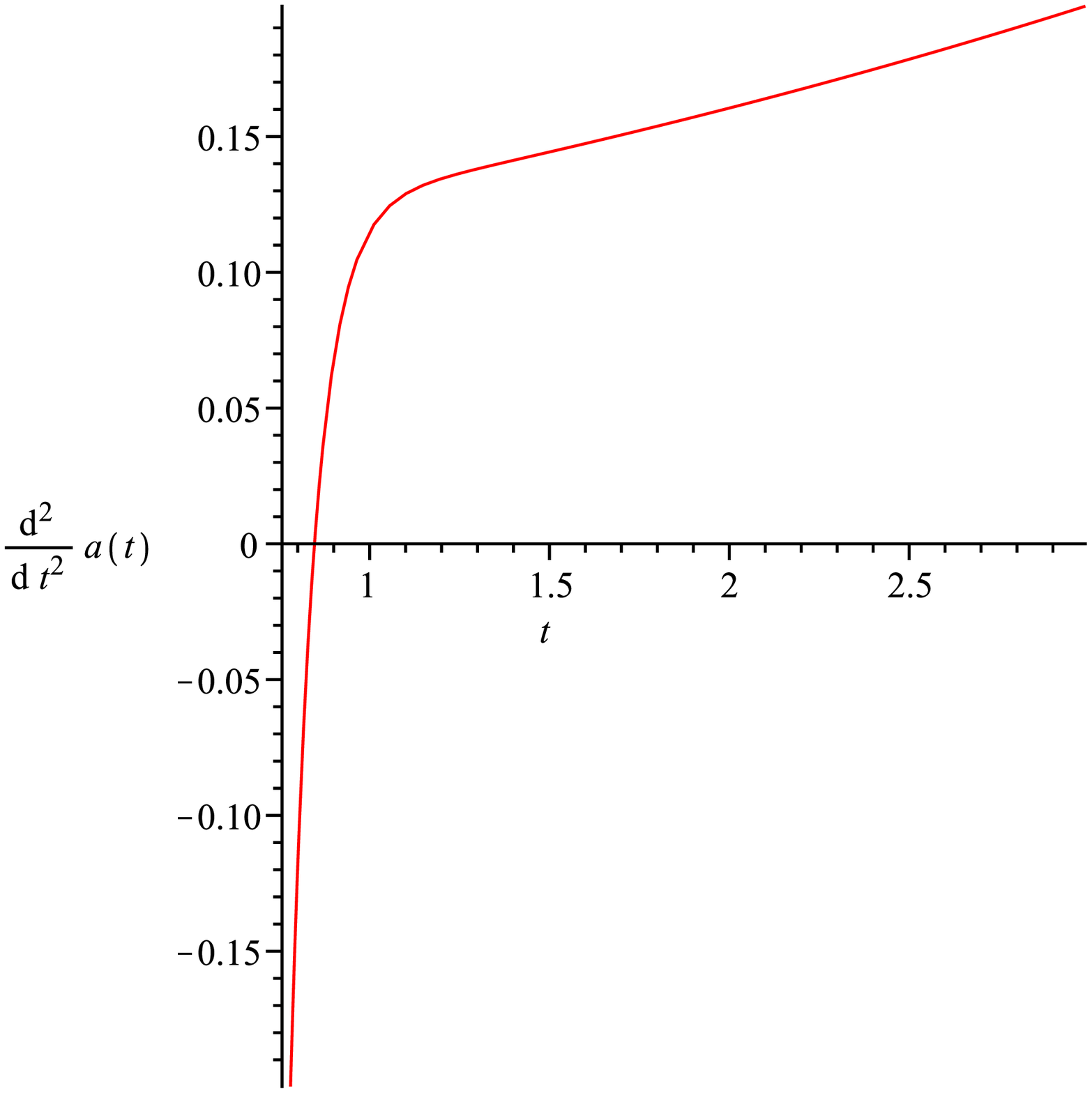}\\
			\caption{variation of acceleration parameter with respect to cosmic time `$t$'~for equation (\ref{k34}) ($\lambda>0$,~ $v_0>0$).}
			\label{fig6}
		\end{minipage}
	\end{figure}

For $\lambda =0$,~~$v_0\neq0$
\begin{eqnarray}
a=e^{\frac{\mu}{\mu-6}\big[\mp\sqrt{|v_0|}(t-t_0)\big]+p_1t+p_2},\nonumber\\
G=e^{\frac{6}{\mu-6}\big[\mp\sqrt{|v_0|}(t-t_0)\big]+p_1t+p_2},\label{k30}
\end{eqnarray}
where $p_1$, $p_2$ are integrating constants\\\\
For $v_0=0,~~\lambda\neq0$
\begin{eqnarray}
 a=e^{m_1t+m_2}\bigg[\mp \sqrt{3|\lambda|}(t-t_0)\bigg]^{-\frac{\mu}{3(\mu-6)}},\nonumber\\
 G=e^{m_1t+m_2}\bigg[\mp \sqrt{3|\lambda|}(t-t_0)\bigg]^{\frac{2}{(\mu-6)}},\label{k31}
\end{eqnarray}
where $m_1$,~~$m_2$ are integrating constants\\\\

For $\lambda<0$,~ $v_0>0$, 
\begin{eqnarray}
a=e^{c_1t+c_2}\bigg[\sqrt{\frac{|\lambda|}{3|v_0|}}sin\bigg(\mp3\sqrt{|v_0|}(t-t_0)\bigg)\bigg]^{-\frac{\mu}{3(\mu-6)}}\nonumber\\
G=e^{c_1t+c_2}\bigg[\sqrt{\frac{|\lambda|}{3|v_0|}}sin\bigg(\mp3\sqrt{|v_0|}(t-t_0)\bigg)\bigg]^{\frac{2}{(\mu-6)}}\label{k32}
\end{eqnarray}
where $c_1$,~$c_2$ are integrating constants\\\\
For $\lambda>0,~~v_0<0$
\begin{eqnarray}
a=e^{c_3t+c_4}\bigg[\sqrt{\frac{|\lambda|}{3|v_0|}}cosh\bigg(\mp3\sqrt{|v_0|}(t-t_0)\bigg)\bigg]^{-\frac{\mu}{3(\mu-6)}}\nonumber\\
G=e^{c_3t+c_4}\bigg[\sqrt{\frac{|\lambda|}{3|v_0|}}cosh\bigg(\mp3\sqrt{|v_0|}(t-t_0)\bigg)\bigg]^{\frac{2}{(\mu-6)}}\label{k33}
\end{eqnarray}
where $c_3$,~~$c_4$ are integrating constants\\\\
For $\lambda>0$,~~$v_0>0$
\begin{eqnarray}
a=e^{c_5t+c_6}\bigg[\sqrt{\frac{|\lambda|}{3|v_0|}}sinh\bigg(\mp3\sqrt{|v_0|}(t-t_0)\bigg)\bigg]^{-\frac{\mu}{3(\mu-6)}}\nonumber\\
G=e^{c_5t+c_6}\bigg[\sqrt{\frac{|\lambda|}{3|v_0|}}sinh\bigg(\mp3\sqrt{|v_0|}(t-t_0)\bigg)\bigg]^{\frac{2}{(\mu-6)}}\label{k34}
\end{eqnarray}
where $c_5$,~$c_6$ are integrating constants.

\section{Cosmological Implications}
Using symmetry analysis, five set of solutions are obtained for the present physical problem for different choices of the parameters involved. To have a clear picture about the possible evolution of the Universe from these solutions graphical presentation of the scale factor (`$a$') and the acceleration parameter (`$\ddot{a}$') are shown in the figures (\ref{fig1}-\ref{fig6}), for the solution sets given by equations (\ref{k32}), (\ref{k33}) and (\ref{k34}) respectively (solutions \ref{k30} and \ref{k31} shown either exponentially expanding or contracting model of the Universe and are not of much interest.) The cosmological solution given by equation (\ref{k32}) represents a bouncing model of the Universe (see figure (\ref{fig1})) where the Universe initially expands and then contracts. Throughout the evolution of the Universe is in decelerating phase (see figure (\ref{fig2})). The solution (\ref{k33}) also represents a bouncing universe but in a reverse way i.e., initially contracting and then expanding (see figure (\ref{fig3})) and the Universe is always in an accelerating phase (see figure (\ref{fig4})). The model of the Universe describing by equation (\ref{k34}) are graphically represented in figures (\ref{fig5}) and (\ref{fig6}). This model is an ever expanding model of the Universe (see figure (\ref{fig5})) and initially the Universe is in a decelerating era and subsequently the Universe will go through an accelerating phase (see figure (\ref{fig6})). So one can speculate that this model represents the Universe from matter dominated era to the present accelerating era of evolution. Therefore, one may conclude that the present variable $G, \Lambda$ theory has possible solution to the present challenging issue of dark energy.  
\section{Acknowledgment}
Author SM thanks CSIR, Govt. of India for awarding Junior research fellowship (File No: 09/096(0892)/2017
). SD acknowledges Science and Engineering Research Board (SERB), Govt. of India, for awarding National Post-
Doctoral Fellowship (File No: PDF/2016/001435) and the Department of Mathematics, Jadavpur University
where a part of the work was completed. SC thanks Science and Engineering Research Board (SERB) for
awarding MATRICS Research Grant support (File No: MTR/2017/000407) and Inter University Center for
Astronomy and Astrophysics (IUCAA), Pune, India for their warm hospitality as a part of the work was done
during a visit.
 

\end{document}